\documentclass{goose-article}

\hypersetup{pdfauthor={%
  T.W.J. de Geus%
}}
\header{%
  T.W.J.~de~Geus, C.~Du, J.P.M.~Hoefnagels, R.H.J.~Peerlings, M.G.D.~Geers\\%
  Scripta Materialia, 2016, 113:101-105, %
  \doi{10.1016/j.scriptamat.2015.10.007}
}

\title{%
  Systematic and objective identification of the microstructure around damage directly from images%
}
\author[1,2]{T.W.J.~de~Geus$^*$}
\author[1,2]{C.~Du}
\author[1]{J.P.M.~Hoefnagels}
\author[1]{R.H.J.~Peerlings}
\author[1]{M.G.D.~Geers}
\affil[1]{%
  Department of Mechanical Engineering, Eindhoven University of Technology, Eindhoven, The Netherlands%
}
\affil[2]{%
  Materials innovation institute (M2i), Delft, The Netherlands%
}
\contact{%
  $^*$Corresponding author:
  \href{mailto:t.w.j.d.geus@tue.nl}{t.w.j.d.geus@tue.nl} %
  \hspace{1mm}--\hspace{1mm} %
  \href{mailto:tom@geus.me}{tom@geus.me} %
  \hspace{1mm}--\hspace{1mm} %
  \href{http://www.geus.me}{www.geus.me}%
}

\begin{document}

\maketitle

\begin{abstract}
An original experimental approach is presented to automatically determine the average phase distribution around damage sites in multi-phase materials. An objective measure is found to be the average intensity around damage sites, calculated using many images. This method has the following benefits: no phase identification or manual interventions are required, and statistical fluctuations and measurement noise are effectively averaged. The method is demonstrated for dual-phase steel, revealing subtle unexpected differences in the morphology surrounding damage in strongly and weakly banded microstructures.
\end{abstract}


\section{Introduction}

Multi-phase materials typically consist of multiple phases with distinct mechanical and physical properties. Their fracture behavior is only partially understood, as the morphology -- often complex -- plays a crucial role (e.g. in multi-phase metals \citep{Tasan2010}, concrete \citep{Elaqra2007}, and geophysics \citep{Torabi2008}). Experimental approaches towards systematic characterization of the microstructural morphology in damaged regions are cumbersome, whereas a reliable methodology might yield new insights and more accurate input for (macroscopic) damage models \citep{Siruguet2004,Hu2007,Maire2008}.

Different statistical descriptors have been developed for arbitrary (microstructural) morphologies. Well known examples are the two-point probability or auto-correlation function and the lineal path function \citep{Torquato1983,Louis1995}. For an isolated inclusion phase (e.g.\ spherical particles) additional descriptors have been developed that convey more information, such as the two-point cluster function and the radial distribution function \citep{Hanish1981}. Almost all measures however require explicit knowledge of the spatial distribution of phases. This knowledge is difficult to obtain experimentally and requires extensive manual processing as the contrast between the phases is often low \citep{Borbely2004}. Furthermore, they are aimed at the quantification of the distribution and/or size of a single phase, while a conditional probability is needed to characterize the neighborhood of a phase (e.g.\ morphology around damage).

In a recent numerical study, De Geus et al.~\citep{DeGeus2015a} characterized the spatial correlation between damage and phase distribution by calculating the average arrangement of phases around damage sites. Extending this analysis to an experimental setting faces the problem that \citep{DeGeus2015a} considered equi-sized grains in the model, corresponding to a finite set of discrete positions (distance measures) that coincide with the grains. In reality the position is continuous (finely discretized experimentally through digital images) and the grains are irregular in position and shape. Furthermore the interpretation in \citep{DeGeus2015a} made use of the explicit knowledge of the phases and damage as a function of the position, not available experimentally.

This letter presents a methodology to quantify the conditional spatial correlation between a uniquely identified feature (e.g.\ damage) and its surrounding morphology directly from a micrograph, without the need for an explicit description of the morphology. As a proof of principle the average arrangement of martensite and ferrite around damage in a dual-phase steel microstructure is characterized. It is well known that in commercial grades martensite often presents a banded structure, which has a strong influence on the damage \citep{Tasan2010}. Two different grades of steel are therefore compared that evidence strongly and weakly banded martensite. Tensile tests on these steel grades show that the weakly banded microstructure has a lower fracture strain, which is in disagreement with the common understanding. The proposed analysis provides novel insights into this topic.

\section{Technique}
\label{sec:technique}

The spatial correlation analysis\footnote{The implementation is open-source. It is optimized, applicable to large sets of high resolution images\\(\href{https://tdegeus@bitbucket.org/tdegeus/gooseeye.git}{https://tdegeus@bitbucket.org/tdegeus/gooseeye.git} and \href{http://www.geus.me/gooseeye}{www.geus.me/gooseeye}).} is discussed in detail in this section, using an artificial example for which the average distribution of two phases around damage sites is quantified based on an image. Several aspects have to be carefully considered to obtain statistically meaningful results. To simplify notation, the analysis is based on fields that are discretized in space.

\begin{figure}[tph]
  \centering
  \includegraphics[width=100mm]{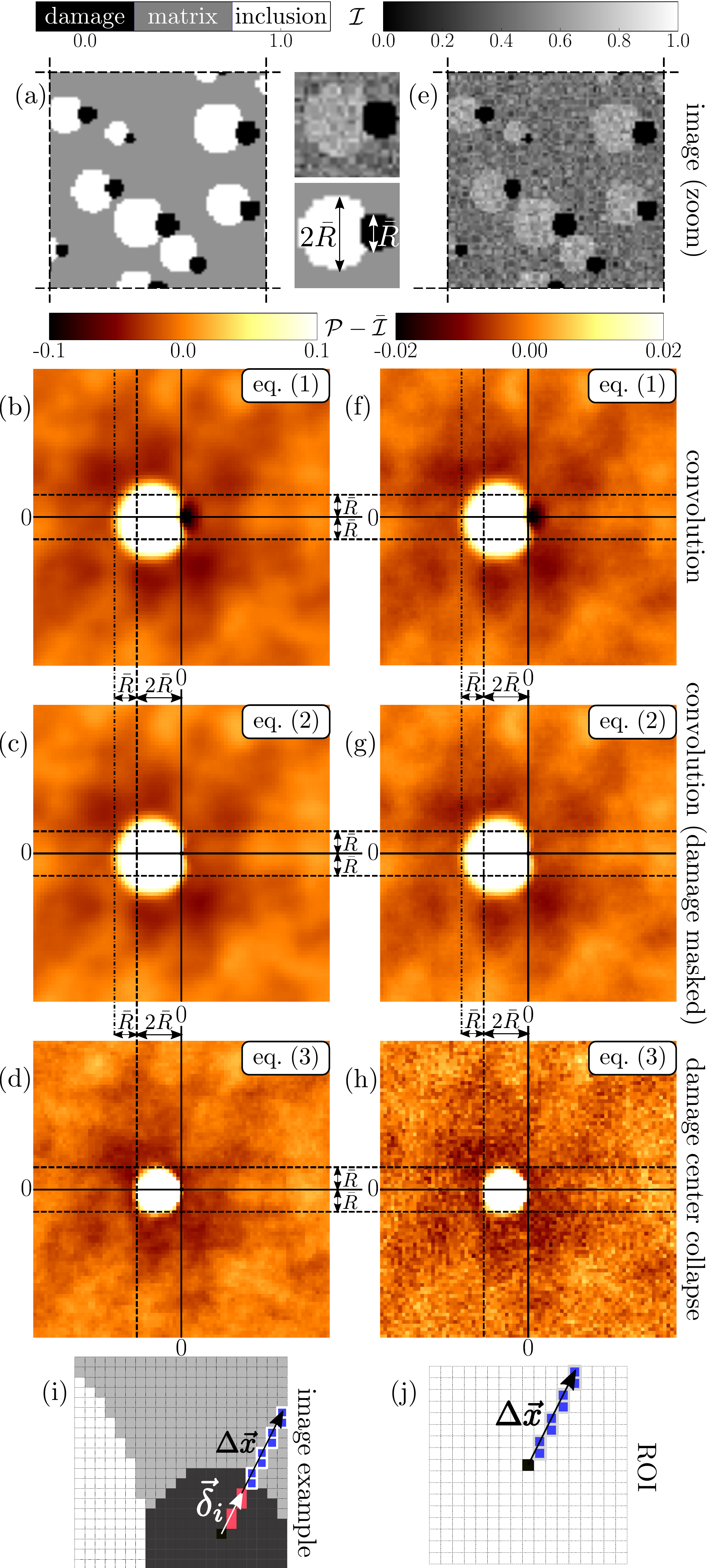}
  \caption{Virtual experiment in the ideal setting: no noise and high phase contrast (a--d), and the realistic setting: with noise and low phase contrast (e--h). From top to bottom: (a,e) the two-phase microstructure, (b--h) the average phase arrangement around a damage site calculated in three different ways. (i--j) An illustration of \eqref{eq:P:collapse-mask} (used in (d,h)).}
  \label{fig:virtual_experiment}
\end{figure}

Consider the example in Figure~\ref{fig:virtual_experiment}(a), which shows part of a periodic microstructure comprising two phases: circular inclusions (white) embedded in a matrix (gray). The inclusions have been numerically generated by randomly perturbing the size and position of an initially regular grid of equi-sized circles with diameter $2 \bar{R}$. Damage (black) is mimicked by shifting each inclusion to the right, applying a position perturbation, and shrinking it by a factor two. These dimensions are indicated in the zoom next to Figure~\ref{fig:virtual_experiment}(a). Two fields are used to describe this image: the image intensity $\mathcal{I}$ and the damage indicator $\mathcal{D}$. For this example $\mathcal{I}(\vec{x}_i) = 1$ in the inclusion phase (white), $\mathcal{I}(\vec{x}_i) = 1/2$ in the matrix (gray), and $\mathcal{I}(\vec{x}_i) = 0$ in damage (black). The damage indicator $\mathcal{D}(\vec{x}_i) = 1$ inside the damage (black) and is zero elsewhere. The position $\vec{x}_i$ denotes the position of a pixel, taken at the position $(i,j)$ in the pixel matrix.

The phase probability $\mathcal{P}$ around damage is calculated as the weighted average
\begin{equation} \label{eq:P-convolution}
  \mathcal{P} (\Delta \vec{x}) =
  \frac{
    \sum_{i}\;
    \mathcal{W} (\vec{x}_i) \;
    \mathcal{I} (\vec{x}_i + \Delta \vec{x})
   }{
     \sum_{i}\;
     \mathcal{W} (\vec{x}_i)
     \hfill
   }
\end{equation}
where the weight factor $\mathcal{W}(\vec{x}_i) = \mathcal{D}(\vec{x}_i)$ for this example. The spatial average is obtained by looping over all pixels $i$ (optionally excluding a boundary region of half the dimensions of the region-of-interest). It thus corresponds to the normalized discrete \textit{convolution} between $\mathcal{W}$ and $\mathcal{I}$. The result is the expectation value of the intensity, $\mathcal{P}$, at a certain position $\Delta \vec{x}$ relative to the damage site. It scales with the image contrast. In the limit case that $\mathcal{I}$ and $\mathcal{W}$ are separate fields that are both explicitly known (i.e.\ zero or one), $\mathcal{P}$ is the probability to find $\mathcal{I}$ at a certain position relative to $\mathcal{W}$.

The analogy of $\mathcal{P}$ with a probability allows the interpretation of its value based on simple statistical arguments. If there is no correlation between $\mathcal{I}$ and $\mathcal{W}$, then $\mathcal{P} = \bar{\mathcal{I}}$, with $\bar{\mathcal{I}}$ the spatial average of $\mathcal{I}$. If, at a position $\Delta \vec{x}$ relative to the damage site, more inclusion phase is found than its spatial average, then $\mathcal{P} (\Delta \vec{x}) > \bar{\mathcal{I}}$ and vice versa.

For the example the result is shown in Figure~\ref{fig:virtual_experiment}(b), where the colormap recovers the extremes (black and white) of the image. Directly to the left of the center (where the damage is) $\mathcal{P} \gg \bar{\mathcal{I}}$, i.e.\ the inclusion phase is identified there. Directly around the center, in all other directions, $\mathcal{P} \approx 0$ which corresponds to damage (black in the image). At larger distance, $\mathcal{P} < \bar{\mathcal{I}}$ corresponding to predominantly matrix phase. Several lighter regions indicate a long-range correlation between damage and inclusion, an intrinsic property of the example for which the inclusion positions are not random but a random perturbation of an initially regular arrangement.

The most obvious artifact in this result is that directly around the damage in the center, damage is identified in a region that corresponds to the size of the damage sites, $\bar{R}$. As the goal is to identify the phase around damage, this cross-correlation of damage should be avoided. It is accounted for through a mask $\mathcal{M}$, which is defined such that $\mathcal{I}(\vec{x}_i)$ is ignored for all pixels where $\mathcal{M}(\vec{x}_i) = 0$. To remove ``damaged'' pixels $\mathcal{M}(\vec{x}_i) = 1 - \mathcal{D}(\vec{x}_i)$. The average phase around damage is now:
\begin{equation} \label{eq:P-convolution-mask}
  \mathcal{P} (\Delta \vec{x}) =
  \frac{
    \sum_{i}\;
    \mathcal{W} (\vec{x}_i) \;
    [ \mathcal{I} \mathcal{M} ] (\vec{x}_i + \Delta \vec{x}) \;
  }{
    \sum_{i}\;
    \mathcal{W} (\vec{x}_i) \;
    \hfill
    \mathcal{M}\, (\vec{x}_i + \Delta \vec{x}) \;
  }
\end{equation}
where the mask in the numerator ensures that the contribution of $\mathcal{I}$ in the damaged areas is omitted, and the mask in the denominator corrects the normalization for the reduced number of data-points. The interpretation of $\mathcal{P}$ is therefore unaffected.

The result is shown in Figure~\ref{fig:virtual_experiment}(c), where the cross-correlation between damage pixels is removed, i.e.\ the black central region in Figure~\ref{fig:virtual_experiment}(b) is absent. Instead, matrix phase is identified there, as expected. Although this result is qualitatively correct, quantitatively the statistical properties of the microstructure have not been preserved. To visualize this, the typical dimensions of the damage and inclusions are indicated in Figure~\ref{fig:virtual_experiment}(c) where the size of the region of elevated inclusion probability (directly to the left of damage) has a diameter of $3 \bar{R}$, while, by statistical arguments, it should have a diameter of $2 \bar{R}$. This results from equation \eqref{eq:P-convolution-mask}, where every damage pixel is separately taken into account. Hence, the resulting phase distribution is smeared over an area equal to the average damage size, in this case $\bar{R}$.

To obtain a more accurate result, the damage site is \textit{collapsed} to a single point by using the analogy to the pore-size probability density (the probability that a point lies at a certain distance of the closest pore-edge \citep{Scheidegger1960}). The basic idea is to quantify the average phase $\mathcal{P}$ at position $\Delta \vec{x}$ relative to the edge of the damage site. Therefore equation~\eqref{eq:P-convolution} is modified to:
\begin{equation} \label{eq:P:collapse-mask}
  \mathcal{P} (\Delta \vec{x}) =
  \frac{
    \sum_{i}\;
    \mathcal{W} (\vec{x}_i) \;
    [ \mathcal{I} \mathcal{M} ] (\vec{x}_i + \vec{\delta}_i(\Delta \vec{x}) +\Delta \vec{x}) \;
  }{
    \sum_{i}\;
    \mathcal{W} (\vec{x}_i) \;
    \hfill
    \mathcal{M}\, (\vec{x}_i + \vec{\delta}_i(\Delta \vec{x})  + \Delta \vec{x}) \;
  }
\end{equation}
where the weight factor $\mathcal{W}(\vec{x}_i)$ equals one only in the geometrical center of the individual damage sites and zero elsewhere, and $\vec{\delta}_i$ is the distance between the damage site center and its edge -- it therefore depends on the orientation of $\Delta \vec{x}$. This is illustrated in Figure~\ref{fig:virtual_experiment}(i), wherein $\vec{\delta}_i$ (red) is that part of the relative position vector inside the damage site, and $\Delta \vec{x}$ (blue) is the part outside the damage site. The resulting $\mathcal{P}$, defined in the region-of-interest (ROI), depends on the distance $\Delta \vec{x}$ only, as shown in Figure~\ref{fig:virtual_experiment}(j). A mask is again used to account for the fact that $\mathcal{I}$ is undefined in the (other) damage sites. Note that the same weight has been assigned to each individual damage site, in different context other choices may be appropriate -- straightforwardly applied to \eqref{eq:P:collapse-mask}.

The result is shown in Figure~\ref{fig:virtual_experiment}(d). As observed, the global pattern is the same as in the earlier results (Figures~\ref{fig:virtual_experiment}(b,c)). The essential difference is that the size of the region of elevated inclusion probability directly left to damage now has diameter $2 \bar{R}$, which coincides with the average inclusion diameter.

Because this method compares the information of all damage sites at once, it is insensitive to (a high degree of) measurement noise and a low intensity contrast between the phases. It only relies on an explicit knowledge of the damage (not of the phases). This is demonstrated by extending the example to a more realistic setting. In Figure~\ref{fig:virtual_experiment}(e), matrix and inclusions are changed to an intensity close to each other and Gaussian noise is added to mimic image noise. The resulting contrast is low, whereby the noise is of the same amplitude as the intensity contrast between the two phases. The results are shown in Figures~\ref{fig:virtual_experiment}(f--h). The orange color is again chosen as the average intensity $\bar{\mathcal{I}}$. The results coincide with the results of Figures~\ref{fig:virtual_experiment}(b--d). A limited amount of noise is still visible in Figure~\ref{fig:virtual_experiment}(h) due to reduced number of data-points (one per damage site). The noise reduces with increasing field-of-view: the root-mean-square of the noise decreases by a factor $\sqrt{n}$ -- with $n$ the number of damage sites. This dependency was verified using the example from this section (results not shown).

\section{Proof of principle: the dual-phase steel case}

\begin{figure}[tph]
  \centering
  \includegraphics[width=105mm]{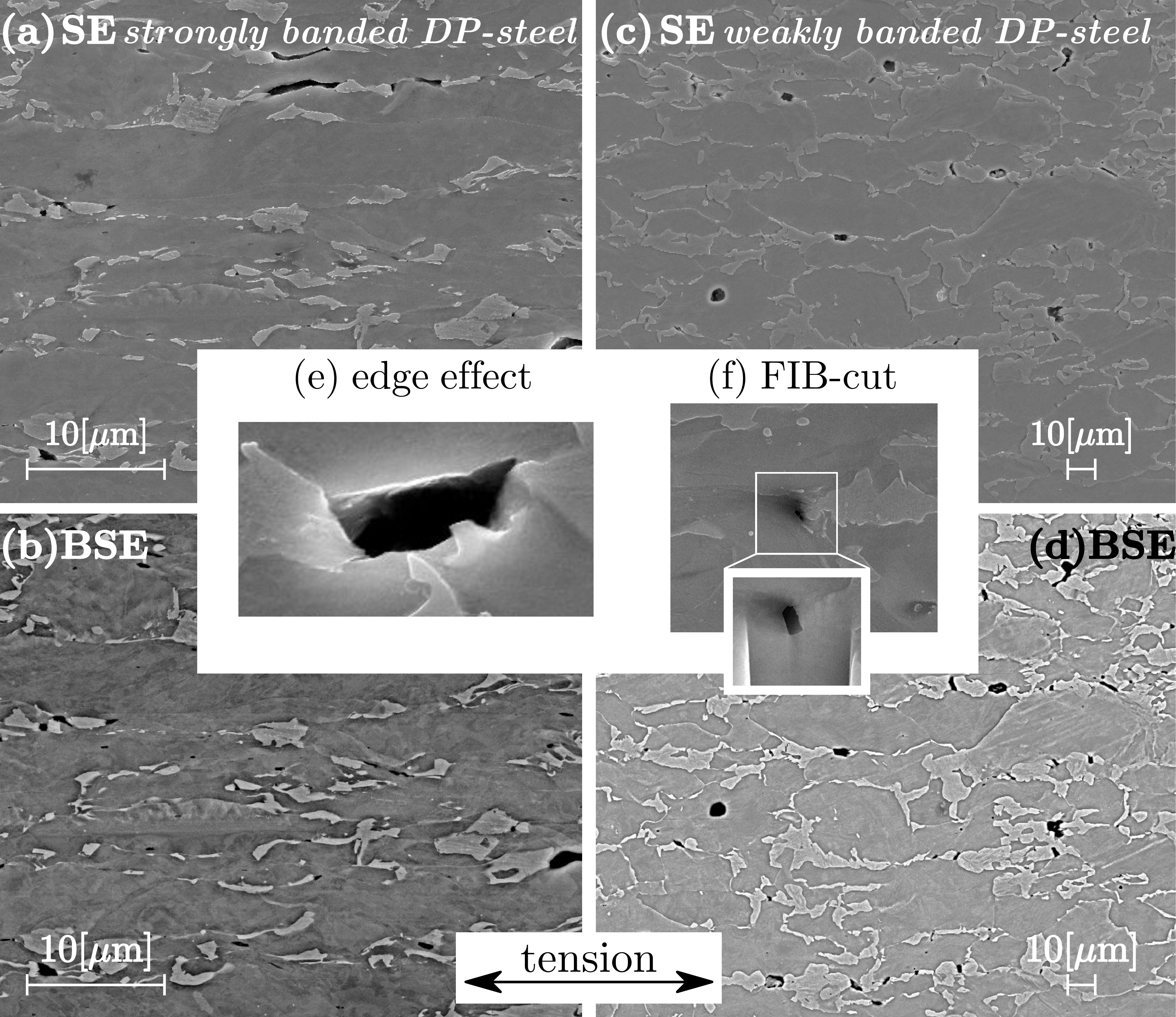}
  \caption{Simultaneously acquired secondary electron (top) and back-scatter electron (bottom) images for strongly (left) and weakly (right) banded dual-phase steel. Example of: (e) the edge-effect related to sharp edges, (f) a focused ion beam milled cross-section of a damage site.}
  \label{fig:typical}
\end{figure}

As case study, the average arrangement of martensite and ferrite around damage in a dual-phase steel is characterized. Two grades are compared: one with strongly banded martensite (commercial DP600) and one which has been heat-treated to remove the martensite bands as much as possible. For both cases, a millimeter-sized tensile specimen is loaded to fracture. The microstructure is examined in the cross-section along the tensile direction, at least $50 \mu \mathrm{m}$ away from the fracture surface. A series of grinding, polishing, and etching steps are applied to create a small height difference between martensite and ferrite, providing contrast in the secondary electron mode of the scanning electron microscope (e.g.\ Figures~\ref{fig:typical}(a,c) for the two grades). In the resulting images martensite is brighter than ferrite. Several damage sites are also visible in Figures~\ref{fig:typical}(a,c), however they cannot be uniquely identified based on intensity alone. To avoid user intervention, a back-scatter electron image is simultaneously acquired to identify the damage uniquely and automatically (see Figure~\ref{fig:typical}(b,d)), as the brightness is zero in the damage sites. This was verified by detailed examination of multiple damaged cross-sections (e.g.\ Figure~\ref{fig:typical}(f)). To establish a statistically representative set, a series of 16 and 11 images were captured of the two grades respectively, whereby all the scan settings were kept constant within each batch of images, resulting in a large field-of-view with high spatial resolution.

\begin{figure}[tph]
  \centering
  \includegraphics[width=95mm]{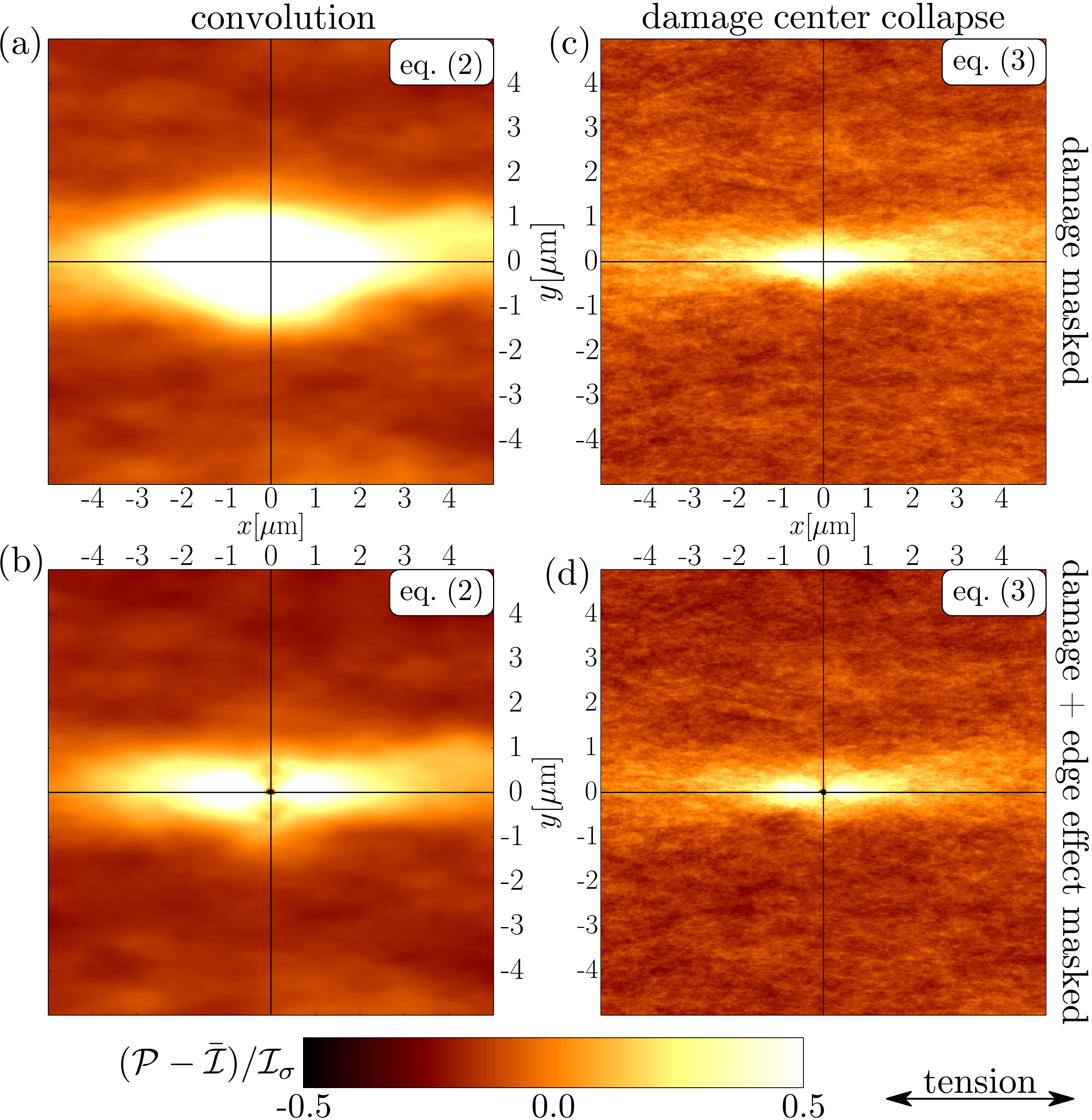}
  \caption{Expectation value of the intensity around damage sites in the strongly banded steel using different approaches (columns) and different masks (rows).}
  \label{fig:hot-spot_calculation}
\end{figure}

The different correlation measures (from Section~\ref{sec:technique}) are compared for the strongly banded dual-phase steel in Figure~\ref{fig:hot-spot_calculation}. Note that the color-scales are normalized with the standard deviation of the intensity, $\mathcal{I}_\sigma$. All results show the same characteristics: damage occurs in a band of martensite aligned with the tensile direction (horizontal) with ferrite in the other directions (top and bottom). However, several correlation measures reveal artifacts for the considered ensemble.

In Figure~\ref{fig:hot-spot_calculation}(a) equation~\eqref{eq:P-convolution-mask} is applied, showing a clear miss-correlation, with $\mathcal{P} \gg \bar{\mathcal{I}}$ in a ring of approximately $2 \mu\mathrm{m}$ around the center. This ring corresponds to the edge-effect around damage (e.g.\ Figure~\ref{fig:typical}(e)), caused by the intrinsic artifact of electron microscopy imaging yielding a strong edge-effect at sharp edges especially in secondary electron mode. This bright ring can be misidentified as martensite and the smearing effect, discussed above, amplifies this artifact.

To resolve the edge-effect, the mask covering each damage site is expanded using standard image dilation. To minimize the loss of information, the number of dilation iterations varies from damage site to damage site and is equal to the square-root of the number of pixels in that damage site. The result is shown in Figure~\ref{fig:hot-spot_calculation}(b), in which the artifact is almost completely removed. What results is the observation that damage occurs in-between regions of martensite that are aligned in the tensile direction with ferrite domains in all other directions.

The applied convolution (equation~\eqref{eq:P-convolution-mask}) has two disadvantages: the result is smeared over a region which scales with the size of the damage, and by definition large damage sites contribute more to the result. In particular the latter may lead to misleading interpretations. To remove this artifact, equation~\eqref{eq:P:collapse-mask} is applied to collapse the damage to a single point in Figures~\ref{fig:hot-spot_calculation}(c--d). As explained, also with the edge-effect unmasked, its influence is substantially reduced as its size is no longer increased during the correlation. The final result, with edge effect masked, is shown in Figure~\ref{fig:hot-spot_calculation}(d). It has the same characteristics as Figure~\ref{fig:hot-spot_calculation}(b), however the regions of martensite are more closely comparable with their average size (estimated from the auto-correlation function, not shown).

\begin{figure}[tph]
  \centering
  \includegraphics[width=105mm]{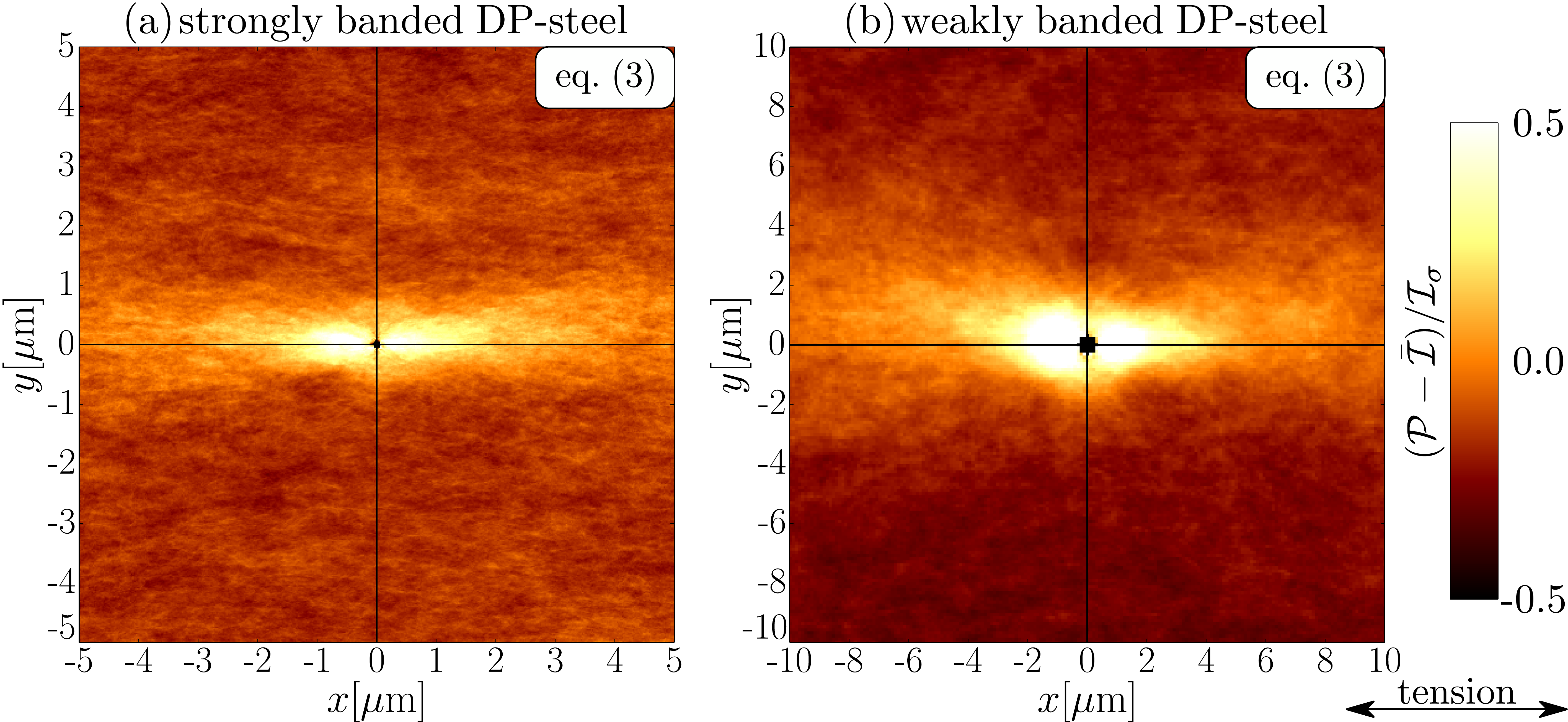}
  \caption{Expectation value of the intensity around damage sites in (a) the strongly and (b) weakly banded dual-phase steel.}
  \label{fig:hot-spot_result}
\end{figure}

Finally, the two different grades -- with strongly and weakly banded martensite -- are compared. The average arrangement of phases around damage for the two grades is shown in Figure~\ref{fig:hot-spot_result}, wherein different axes are used to correct for the different grain sizes of the grades. The results have the same overall pattern: damage occurs in-between martensite aligned with the tensile direction with ferrite in the other directions. For the strongly banded microstructure, the martensite appears in bands (Figure~\ref{fig:hot-spot_result}(a)), whereas for the weakly banded microstructure the martensite confined in a relatively small region (Figure~\ref{fig:hot-spot_result}(b)). This implies that even if the bands are not present, damage still occurs in-between martensite domains.

The two grades of steel are quantitatively compared to reveal a surprising difference. Whereby the scaling of expectation value, $\mathcal{P}$, with the image contrast, $\mathcal{I}_\sigma$, is employed to obtain a probability measure that is independent of the image contrast. For Figure~\ref{fig:hot-spot_result} this implies that the difference in color between the two results may be interpreted as a difference in martensite and ferrite probability around the damage sites. The darker regions above and below the damage sites in Figure~\ref{fig:hot-spot_result}(b) compared with Figure~\ref{fig:hot-spot_result}(a) indicate that the probability of ferrite is lower for the strongly banded microstructure, i.e.\ the martensite bands are located in clusters. This gives rise to a hypothesis: as the fracture strain is 6\% higher for the strongly banded microstructure, the presence of clusters of martensite above and below the damage may delay propagation. Revisiting the images, e.g.\ Figure~\ref{fig:typical}, confirms this observation. Although further analysis is needed, it is interesting to see how new insights and hypotheses can originate from the presented, objective, analysis.

\section{Conclusion and outlook}

A methodology has been presented to quantify the spatial arrangement of phases around damage sites directly using images. This technique is quite general: it may correlate different microstructural features such as phases, cavities, etc.\ in the fields of materials science, geophysics, medicine and many others. It requires no phase identification, no user interaction, and automatically averages out noise. The interpretation of the result is straightforward as it retains the properties of the original images. In the context of materials, this technique can readily be applied to three-dimensional tomographic or serial sectioning measurements, e.g.\ from \citep[][]{Borbely2004,Li1999,Landron2012}, for which it is nearly impossible to perform systematic identifications without an automated analysis technique. Finally the approach does not rely on a high contrast, which is usually hard to obtain experimentally.

A case study resulted in the average arrangement of martensite and ferrite around damage in the microstructure of a dual-phase steel. This has led to a surprising observation, easily overlooked otherwise. This is the key added value of this methodology.

The results also open up the possibility to use the identified average arrangement of phases as a predictive tool. Indeed in a preliminary study, using the numerical results of \citep{DeGeus2015a}, the fracture initiation sites have been identified accurately.

\section*{Acknowledgments}

This research was carried out under project number M22.2.11424 in the framework of the research program of the Materials innovation institute M2i (\href{http://www.m2i.nl}{www.m2i.nl}).

\section*{References}

\scriptsize
\bibliography{library}

\begin{thebibliography}{14}
\providecommand{\natexlab}[1]{#1}
\providecommand{\url}[1]{\texttt{#1}}
\expandafter\ifx\csname urlstyle\endcsname\relax
  \providecommand{\eprint}[1]{eprint: #1}\else
  \providecommand{\eprint}{eprint: \begingroup \urlstyle{rm}\Url}\fi
\expandafter\ifx\csname urlstyle\endcsname\relax
  \providecommand{\doi}[1]{doi: #1}\else
  \providecommand{\doi}{doi: \begingroup \urlstyle{rm}\Url}\fi

\bibitem[Tasan et~al.(2010)Tasan, Hoefnagels, and Geers]{Tasan2010}
C.C.\ Tasan, J.P.M.\ Hoefnagels, and M.G.D.\ Geers.
\newblock {Microstructural banding effects clarified through micrographic
  digital image correlation}.
\newblock \emph{Scr.\ Mater.}, 62\penalty0 (11):\penalty0 835--838, 2010.
\newblock \doi{10.1016/j.scriptamat.2010.02.014}.

\bibitem[Elaqra et~al.(2007)Elaqra, Godin, Peix, R'Mili, and
  Fantozzi]{Elaqra2007}
H.\ Elaqra, N.\ Godin, G.\ Peix, M.\ R'Mili, and G.\ Fantozzi.
\newblock {Damage evolution analysis in mortar, during compressive loading
  using acoustic emission and X-ray tomography: Effects of the sand/cement
  ratio}.
\newblock \emph{Cem.\ Concr.\ Res.}, 37\penalty0 (5):\penalty0 703--713, 2007.
\newblock \doi{10.1016/j.cemconres.2007.02.008}.

\bibitem[Torabi et~al.(2008)Torabi, Fossen, and Alaei]{Torabi2008}
A.\ Torabi, H.\ Fossen, and B.\ Alaei.
\newblock {Application of spatial correlation functions in permeability
  estimation of deformation bands in porous rocks}.
\newblock \emph{J.\ Geophys.\ Res.\ Solid Earth}, 113\penalty0 (8):\penalty0
  1--10, 2008.
\newblock \doi{10.1029/2007JB005455}.

\bibitem[Siruguet and Leblond(2004)]{Siruguet2004}
K.\ Siruguet and J.-B.\ Leblond.
\newblock {Effect of void locking by inclusions upon the plastic behavior of
  porous ductile solids—I: theoretical modeling and numerical study of void
  growth}.
\newblock \emph{Int.\ J.\ Plast.}, 20\penalty0 (2):\penalty0 225--254, 2004.
\newblock \doi{10.1016/S0749-6419(03)00018-4}.

\bibitem[Hu et~al.(2007)Hu, Bai, and Ghosh]{Hu2007}
C.\ Hu, J.\ Bai, and S.\ Ghosh.
\newblock {Micromechanical and macroscopic models of ductile fracture in
  particle reinforced metallic materials}.
\newblock \emph{Model.\ Simul.\ Mater.\ Sci.\ Eng.}, 15\penalty0 (4):\penalty0
  S377--S392, 2007.
\newblock \doi{10.1088/0965-0393/15/4/S05}.

\bibitem[Maire et~al.(2008)Maire, Bouaziz, {Di Michiel}, and Verdu]{Maire2008}
E.\ Maire, O.\ Bouaziz, M.\ {Di Michiel}, and C.\ Verdu.
\newblock {Initiation and growth of damage in a dual-phase steel observed by
  X-ray microtomography}.
\newblock \emph{Acta Mater.}, 56\penalty0 (18):\penalty0 4954--4964, 2008.
\newblock \doi{10.1016/j.actamat.2008.06.015}.

\bibitem[Torquato and Stell(1983)]{Torquato1983}
S.\ Torquato and G.\ Stell.
\newblock {Microstructure of two-phase random media. III. The n-point matrix
  probability functions for fully penetrable spheres}.
\newblock \emph{J.\ Chem.\ Phys.}, 79\penalty0 (3):\penalty0 1505, 1983.
\newblock \doi{10.1063/1.445941}.

\bibitem[Louis and Gokhale(1995)]{Louis1995}
P.\ Louis and A.M.\ Gokhale.
\newblock {Application of Image Analysis for Characterization of Spatial
  Arrangements of Features in Microstructure}.
\newblock \emph{Metall.\ Mater.\ Trans.\ A}, 26:\penalty0 1449--1456, 1995.
\newblock \doi{10.1007/BF02647595}.

\bibitem[Hanish and Stoyan(1981)]{Hanish1981}
K.-H.\ Hanish and D.\ Stoyan.
\newblock {Stereological estimation of the radial distribution function of
  centres of spheres}.
\newblock \emph{J.\ Microsc.}, 122\penalty0 (2):\penalty0 131--141, 1981.
\newblock \doi{10.1111/j.1365-2818.1981.tb01252.x}.

\bibitem[Borb{\'{e}}ly et~al.(2004)Borb{\'{e}}ly, Csikor, Zabler, Cloetens, and
  Biermann]{Borbely2004}
A.\ Borb{\'{e}}ly, F.F.\ Csikor, S.\ Zabler, P.\ Cloetens, and H.\ Biermann.
\newblock {Three-dimensional characterization of the microstructure of a
  metal-matrix composite by holotomography}.
\newblock \emph{Mater.\ Sci.\ Eng.\ A}, 367\penalty0 (1-2):\penalty0 40--50,
  2004.
\newblock \doi{10.1016/j.msea.2003.09.068}.

\bibitem[de~Geus et~al.(2015)de~Geus, Peerlings, and Geers]{DeGeus2015a}
T.W.J.\ de~Geus, R.H.J.\ Peerlings, and M.G.D.\ Geers.
\newblock {Microstructural topology effects on the onset of ductile failure in
  multi-phase materials - A systematic computational approach}.
\newblock \emph{Int.\ J.\ Solids Struct.}, 67-68:\penalty0 326--339, 2015.
\newblock \doi{10.1016/j.ijsolstr.2015.04.035}.
\newblock \eprint{1604.02858}.

\bibitem[Scheidegger(1960)]{Scheidegger1960}
A.E.\ Scheidegger.
\newblock \emph{{The physics of flow through porous media}}.
\newblock Macmillan, 1960.

\bibitem[Li et~al.(1999)Li, Ghosh, and Richmond]{Li1999}
M.\ Li, S.\ Ghosh, and O.\ Richmond.
\newblock {An experimental-computational approach to the investigation of
  damage evolution in discontinuously reinforced aluminum matrix composite}.
\newblock \emph{Acta Mater.}, 47\penalty0 (12):\penalty0 3515--3532, 1999.
\newblock \doi{10.1016/S1359-6454(99)00148-2}.

\bibitem[Landron et~al.(2012)Landron, Maire, Adrien, Suhonen, Cloetens, and
  Bouaziz]{Landron2012}
C.\ Landron, E.\ Maire, J.\ Adrien, H.\ Suhonen, P.\ Cloetens, and O.\ Bouaziz.
\newblock {Non-destructive 3-D reconstruction of the martensitic phase in a
  dual-phase steel using synchrotron holotomography}.
\newblock \emph{Scr.\ Mater.}, 66\penalty0 (12):\penalty0 1077--1080, 2012.
\newblock \doi{10.1016/j.scriptamat.2012.03.003}.

\end{thebibliography}

\end{document}